\documentclass[a4paper]{jpconf}

\usepackage{amsmath}
\usepackage{amsthm}
\usepackage{amssymb}
\usepackage{wasysym}
\usepackage{graphicx}%
\usepackage{bm}%
\usepackage{bbold}%
\usepackage{xcolor}%
\usepackage[caption=false]{subfig}%
\usepackage{cite}

\usepackage[pdftex]{hyperref} %
\definecolor{darkblue}{rgb}{0,0,.5} %
\definecolor{black}{rgb}{0,0,0} %
\hypersetup{
  colorlinks=true, breaklinks=true, linkcolor=darkblue,
  menucolor=darkblue,
  urlcolor=darkblue, citecolor=darkblue }

\def\tr{\operatorname{tr}}
\def\Tr{\operatorname{Tr}}

\def\mc{\mathcal}
\def\bs{\boldsymbol}

\def\op{}
\def\mat{\bs}
\def\fcl{\widehat}
\def\gop{\mc}

\def\ord{\mc T}

\def\cre#1#2{#1^\dagger_{#2}}%
\def\ann#1#2{#1^{\vphantom{\dagger}}_{#2}}%

\def\cc#1{\cre{c}{#1}}%
\def\ac#1{\ann{c}{#1}}%
\def\n#1{n_{#1}}%

\def\ra{\rightarrow}%
\def\ua{\uparrow}%
\def\da{\downarrow}%

\def\com[#1,#2]{\left[#1,#2\right]}
\def\contcom[#1,#2]{\left[#1\stackrel{\circ}{,}#2\right]}

\def\ev<#1>{\left<#1\right>}

\def\bra<#1|{\left<#1\right|}
\def\ket|#1>{\left|#1\right>}
\def\braket<#1|#2|#3>{\langle#1|#2|#3\rangle}

\def\dint{\int\negthickspace\negthickspace\int}

\begin{document}
  
\title{Nonequilibrium variational-cluster approach to real-time
  dynamics in the Fermi-Hubbard model}

\author{Felix Hofmann$^1$, Martin Eckstein$^2$ and Michael
  Potthoff$^1$} \address{$^1$ I. Institut f\"ur Theoretische Physik,
  Universit\"at Hamburg, Jungiusstra\ss{}e 9, 20355 Hamburg, Germany}
\address{$^2$ Max Planck Research Department for Structural Dynamics
  at the University of Hamburg, CFEL, Notkestra\ss{}e 85, 22607
  Hamburg, Germany} \ead{fhofmann@physik.uni-hamburg.de}

\begin{abstract}
  The nonequilibrium variational-cluster approach is applied to study
  the real-time dynamics of the double occupancy in the
  one-dimensional Fermi-Hubbard model after different fast changes of
  hopping parameters.  A simple reference system, consisting of
  isolated Hubbard dimers, is used to discuss different aspects of the
  numerical implementation of the approach in the general framework of
  nonequilibrium self-energy functional theory.  Opposed to a direct
  solution of the Euler equation, its time derivative is found to
  serve as numerically tractable and stable conditional equation to
  fix the time-dependent variational parameters.
\end{abstract}

\section{Introduction}
\label{sec:introduction}

Over the past decade, considerable progress in the field of ultracold
gases has given access to experimentally simulating prototypical
many-body models, e.g., known from condensed-matter physics, with a
high degree of dynamic control
\cite{bloch2008,lewenstein2007,dutta2015}.  Accompanied by theoretical
work, this opened up the possibility to study their emergent phases
and nonequilibrium dynamics induced by fast changes of the model
parameters \cite{polkovnikov2011,eisert2015}.  As \emph{the}
paradigmatic model for strongly correlated fermions on a lattice, we
consider the Fermi-Hubbard model
\cite{gutzwiller1963,hubbard1963,kanamori1963}.  Not only the bosonic
variant \cite{fisher1989,jaksch1998,greiner2002,krutitsky2015} but
also the Fermi-Hubbard model has been realized by recent
optical-lattice experiments
\cite{koehl2005,giorgini2008,esslinger2010}. This promises to provide
insight into fundamental questions related to the Mott transition, to
collective order or even to non-Fermi-liquid physics.

Using standard notations, the Fermi-Hubbard Hamiltonian is given by
\begin{equation}
  \label{eq:Hamiltonian}
  \op H(t) = \sum_{\langle ij \rangle,\sigma}T_{ij}(t)\cc{i\sigma}\ac{j\sigma} +
  U(t) \sum_{i} \n{i\ua}\n{i\da} \,,
\end{equation}
where a fermion at site $i$ and with spin projection $\sigma=\uparrow,
\downarrow$ is annihilated (created) by $c^{(\dagger)}_{i\sigma}$, and
where $\n{i\sigma} = \cc{i\sigma}\ac{i\sigma}$ is the density
operator.  Tunneling between neighboring sites $\langle ij \rangle$ is
described by the hopping amplitude $T_{ij}(t)$.  On the same site,
fermions are subjected to the repulsive Hubbard interaction $U(t)$.  A
nonequilibrium state and nontrivial real-time dynamics of
observables can be initiated by fast changes of the time-dependent
model parameters $T_{ij}(t)$ and $U(t)$.
 
Despite its conceptual simplicity, the Hubbard model has challenged
theoreticians for more than five decades, and has stimulated the
development of a large number of different methods.  Many approaches
which are based on the language of Green's functions \cite{KMS}
have successfully been extended to describe real-time phenomena
\cite{thygesen2007,schmidt2002,freericks2006,aoki2013,munoz2013,jung2012,knap2011,balzer2013,lipavsky1986},
using the Keldysh formalism \cite{keldysh1965}.  Of particular
interest in this context are quantum-cluster theories
\cite{maier2005}, which are formulated in the thermodynamic limit, but
nevertheless account for nonlocal correlations on a length scale
defined by the linear extension of a reference cluster, which can have
a profound influence on the nonequilibrium dynamics
\cite{tsuji2013,eckstein2014}.  A simple but conceptually appealing
quantum-cluster approach is the cluster-perturbation theory (CPT)
\cite{gros1993,senechal2000,potthoff2011c,gramsch2015}.  Within the
CPT, the infinite lattice is tiled into small clusters, which can be
solved exactly by numerical means.  This solution is then extended to
the full lattice by infinite-order perturbation theory with respect to
the inter-cluster hopping but neglecting vertex corrections
\cite{potthoff2011c}.  This amounts to approximate the self-energy of
the full model by the self-energy of the reference system of
disconnected clusters.  A major drawback of the CPT consists in the
fact that, even for a given geometrical tiling of the lattice, the
partitioning of the hopping part and hence the choice of the reference
system, i.e., the starting point of the perturbative expansion is not
unique.  However, the nonuniqueness of the CPT construction can be
turned into an advantage, if one can find a variational prescription
for finding the \emph{optimal} starting point for the
cluster-perturbation theory.  In fact, this is achieved with the
variational-cluster approach (VCA) \cite{dahnken2004,potthoff2014}
which is best understood in the general framework of self-energy
functional theory \cite{potthoff2003,hofmann2013}.

The main purpose of the present paper is to discuss different
practical issues related to the application of the {\em
  nonequilibrium} VCA.  While the main theoretical concept is
essentially the same as for the equilibrium VCA, we demonstrate that
the numerical implementation of the nonequilibrium variant of the
approach is by far more complex and requires new techniques.  As a
proof of principle, we present a first numerical implementation of the
nonequilibrium VCA, based on simple two-site reference clusters, and
study two types of parameter quenches (or fast ramps) in the
one-dimensional Fermi-Hubbard model with alternating (dimerized)
hopping amplitudes which have attracted experimental attention
recently \cite{greif2015,atala2013}.

The paper is organized as follows: In
Section~\ref{sec:noneq-greens-funct} we briefly review the concept of
nonequilibrium Green's functions used for the nonequilibrium
extension of the self-energy functional theory, as described in
Sec.~\ref{sec:self-energy-funct}.  We derive the central Euler
equation in Sec.~\ref{sec:euler} and discuss its numerical
implementation in Sec.~\ref{sec:numer-impl}.  Results for the
dimerized Hubbard model are presented in Sec.~\ref{sec:nevca}.  A
summary is given in Sec.~\ref{sec:summary-outlook}.


\section{Nonequilibrium Green's functions}
\label{sec:noneq-greens-funct}

The self-energy functional theory relies on functionals that are
formally defined by means of all-order perturbation theory and thus on
the concept of (nonequilibrium) Green's functions
\cite{KMS,keldysh1965}.
Throughout this paper, we make use of the general theory provided by
Refs.~\cite{DLRK} 
but in first place closely follow the formal setup by Wagner
\cite{wagner1991}.

\begin{figure}[t]
  \centering
  \includegraphics[width=.58\textwidth]{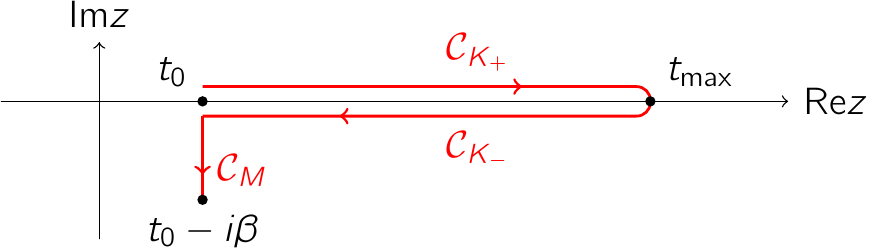}\hspace{.04\textwidth}%
  \begin{minipage}[b]{.38\textwidth}
    \caption{The three-branched Keldysh-Matsubara contour $\mc C$ in
      the complex time plane, extending up to time $t_{\rm max}$; $\mc
      C_{K_\pm}$ denotes the upper/lower branch and $\mc C_M$ the
      Matsubara branch.}
    \label{fig:contour}
  \end{minipage}
\end{figure}

For a general fermionic lattice model with Hamiltonian
\begin{equation}
  \label{eq:generalHamiltonian}
  \op H_{\mat T,\mat U} (t) = \sum_{\alpha\beta}T_{\alpha\beta}(t)\cc{\alpha}\ac{\beta} +
  \frac{1}{2}\sum_{\alpha\beta\gamma\delta}
  U_{\alpha\beta\delta\gamma}(t)\cc{\alpha}\cc{\beta}\ac{\gamma}\ac{\delta} \, ,
\end{equation}
which is specified by the parameters $\mat T$ and $\mat U$, and which
initially (at time $t=t_{0}$) is prepared in a thermal state with
inverse temperature $\beta$ and chemical potential $\mu$, the elements
of the contour-ordered Green's function $\mat G_{\mat T,\mat U}$ are
defined as
\begin{equation}
  \label{eq:DefNEGF}
  i G_{\mat T,\mat U}(1,2) = \ev<\ord_{\mc C} \ac{\gop H}(1)\cc{\gop H}(2) >_{\mat T, \mat U} \, .
\end{equation}
Here, greek indices combine all one-particle orbitals, and
$c^{(\dagger)}_{\gop H}(i)$ is the annihilator (creator) in the
Heisenberg picture with respect to $\gop H(z) = \op H(z) - \mu N$,
where $N$ is the total particle-number operator.  We use the
short-hand notation $i \equiv (\alpha_i,z_i)$, where $z_i$ marks an
arbitrary point on the Keldysh-Matsubara contour $\mc C$ (see
Fig.~\ref{fig:contour}).  $\ord_{\mc C}$ is the contour time ordering.
Further, $\ev<\cdots> = \tr(\rho\,\cdots)$ denotes the expectation
value in the initial state, where $\rho$ is density operator, $\rho =
\exp(-\beta \gop H_{\rm ini})/Z$, and $Z = \tr \exp(-\beta \gop H_{\rm
  ini})$ with $\gop H_{\rm ini} = \op H(t_0) - \mu \op N$ is the
partition function.

From the Heisenberg equation of motion for the annihilator, we find
the equation of motion
\begin{equation}
  \left( i \partial_z - (\mat T(z) - \mu) \right) \mat
  G_{\mat T,\mat U}(z,z') = \mat 1 \delta_{\mc C}(z,z') + \left(
    \mat \Sigma_{\mat T,\mat U} \circ \mat G_{\mat T,\mat U}
  \right)(z,z') \, ,  
  \label{eq:EOM-NEGF}
\end{equation}
where $\delta_{\mc C}$ is the contour delta-function and
$\mat\Sigma_{\mat T,\mat U}$ is the self-energy. The circle $\circ$ on
the right-hand side stands for an integration along $\mc C$ and an
implicit summation over all orbital indices. By setting $\mat U=0$ in
Eq.~(\ref{eq:EOM-NEGF}), we obtain the inverse of the ``free'' Green's
function $\mat G_{\mat T,0}$ as
\begin{equation}
  \label{eq:inverseFreeNEGF}
  G^{-1}_{\mat T,0;\alpha\alpha'}(z,z') = \delta_{\mc C}(z,z') \left(
    \delta_{\alpha\alpha'} i \partial_{z'} -
    \left(T_{\alpha\alpha'}(z') - \mu
      \delta_{\alpha\alpha'}\right) \right) \, .
\end{equation}
Using this, we can rewrite Eq.~(\ref{eq:EOM-NEGF}) as Dyson's
equation:
\begin{equation}
  \label{eq:DysonEqnShort}
  \mat G_{\mat T,\mat U} = \mat G_{\mat T,0} + \mat G_{\mat
    T,0} \circ \mat \Sigma_{\mat T,\mat U} \circ \mat G_{\mat T,\mat
    U} \, . 
\end{equation}


\section{Self-energy functional theory}
\label{sec:self-energy-funct}

The (nonequilibrium) self-energy functional theory is based on the
following functional of the nonequilibrium self-energy (see
Refs.~\cite{potthoff2006b,hofmann2013} for details):
\begin{equation}
  \label{eq:NESEfcl}
  \fcl\Omega_{\mat T,\mat U}[\mat\Sigma] 
  = 
  \frac{1}{\beta} \Tr\ln
  \left( \mat G_{\mat T,0}^{-1} - \mat\Sigma \right)^{-1} 
  + 
  \fcl
  F_{\mat U} [\mat \Sigma] \, .
\end{equation}
Here, the first term on the right-hand side is a simple and explicit
functional of $\mat \Sigma$ and depends on the system's one-particle
parameters $\mat T$.  The second term is the Legendre transform of the
Luttinger-Ward functional $\fcl\Phi_{\mat U}[\mat G]$
\cite{luttinger1960}.  The latter can be defined by means of all-order
perturbation theory or, nonperturbatively, within a path-integral
formalism analogous to the equilibrium case \cite{potthoff2006b}.
Note, that functionals are indicated by a hat, and that the trace is
defined as $\Tr \mat A = \sum_\alpha \int_{\mc C} d z\,
A_{\alpha\alpha}(z,z^+)$, where $z^+$ is infinitesimally later than
$z$ on $\mc C$.

There are two important properties of the self-energy functional
(\ref{eq:NESEfcl}): First, if $\fcl\Omega_{\mat T,\mat U}[\mat\Sigma]$
is evaluated at the physical self-energy $\mat \Sigma = \mat
\Sigma_{\mat T, \mat U}$, one obtains the physical grand potential of
the system in its initial thermal state.  All contributions from the
upper and the lower Keldysh branch cancel in this case.  Second, the
self-energy functional is stationary at the physical self-energy,
i.e.,
\begin{equation}
  \left.\frac{\delta \fcl{\Omega}_{\mat T, \mat U}[\mat \Sigma]}
    {\delta \mat \Sigma}\right|_{\mat \Sigma = \mat \Sigma_{\mat T, \mat
      U}} 
  = 
  0 \, .
  \label{eq:StatPointNESEfcl}
\end{equation}
In fact, using the well-known properties of the Luttinger-Ward
functional and its Legendre transform, it is easy to see that Eq.\
(\ref{eq:StatPointNESEfcl}) is equivalent with Dyson's
equation~(\ref{eq:DysonEqnShort}).  Eq.\ (\ref{eq:StatPointNESEfcl})
constitutes a general variational principle by which one could
determine the self-energy (as well as the grand potential) of a system
of correlated fermions.  However, the explicit form of the functional
$\fcl F_{\mat U} [\mat \Sigma]$ is generally unknown.

An important observation is that the Luttinger-Ward functional, and
thus also $\fcl F_{\mat U} [\mat \Sigma]$ is universal, i.e., it
depends (besides the argument) on the interaction parameters $\mat U$
only, as made explicit by the subscript.  Hence, the functional form
of $\fcl F_{\mat U} [\mat \Sigma]$ remains unchanged for any reference
system with the same interaction $\mat U$ as the original system but
with a different set of one-particle parameters $\mat\lambda'$, i.e.,
for any reference system with a Hamiltonian of the form $\op H' \equiv
\op H_{\mat \lambda',\mat U}$.

Let us choose a reference system in a particular subspace of
one-particle parameters $\mat \lambda'$, which has a sufficiently
simple structure so that its one-particle Green's function, its
self-energy, and its grand potential of the initial thermal state at
$\beta$ and $\mu$ can be computed exactly for any $\mat \lambda'$ of
the subspace.  (Here and in the following, primed quantities will
refer to the reference system.)  We can write down the self-energy
functional of the reference system,
\begin{equation}
  \label{eq:NESEfclRef}
  \fcl\Omega_{\mat \lambda',\mat U}[\mat\Sigma] 
  = 
  \frac{1}{\beta}
  \Tr\ln \left( \mat G_{\mat \lambda',0}^{-1} - \mat\Sigma
  \right)^{-1} 
  + 
  \fcl F_{\mat U} [\mat \Sigma] \, ,
\end{equation}
and, due to the mentioned universality, eliminate $\fcl F_{\mat
  U}[\mat\Sigma]$ by combining Eqs.~(\ref{eq:NESEfcl}) and
(\ref{eq:NESEfclRef}):
\begin{equation}
  \label{eq:NESEfclDiff}
  \fcl\Omega_{\mat T,\mat U}[\mat\Sigma] 
  = 
  \fcl\Omega_{\mat
    \lambda',\mat U}[\mat\Sigma] + \frac{1}{\beta} \Tr\ln
  \left( \mat G_{\mat T,0}^{-1} - \mat\Sigma \right)^{-1} 
  - 
  \frac{1}{\beta} \Tr\ln \left( \mat G_{\mat \lambda',0}^{-1} -
    \mat\Sigma \right)^{-1} \,.
\end{equation}
This expression for the self-energy functional, which is still exact,
can be evaluated in practice for a certain subclass of trial
self-energies, namely for the physical self-energies of the reference
system.  Inserting $\mat \Sigma = \mat \Sigma_{\mat \lambda', \mat
  U}$, we have
\begin{equation}
  \label{eq:SFTfcl}
  \fcl\Omega_{\mat T,\mat U}[\mat\Sigma_{\mat \lambda',\mat U}] 
  =
  \Omega_{\mat \lambda',\mat U} + \frac{1}{\beta} \Tr\ln \left( \mat
    G_{\mat T,0}^{-1} - \mat\Sigma_{\mat \lambda',\mat U} \right)^{-1}
  - 
  \frac{1}{\beta} \Tr\ln \left( \mat G_{\mat \lambda',\mat U}
  \right) \,,
\end{equation}
where we used $\fcl\Omega_{\mat \lambda', \mat U}[\mat \Sigma_{\mat
  \lambda',\mat U}]= \Omega_{\mat \lambda',\mat U}$ and Dyson's
equation for the reference system, see Eq.\ (\ref{eq:DysonEqnShort}).
Using the stationarity principle Eq.\ (\ref{eq:StatPointNESEfcl}), we
can optimize the self-energy over the restricted subspace of trial
self-energies spanned by the subspace of variational parameters $\mat
\lambda'$ via
\begin{equation}
  \label{eq:EulerEqn}
  \left. \frac{\delta \fcl\Omega_{\mat T,\mat U}[\mat\Sigma_{\mat \lambda',\mat U}]}{\delta \mat \lambda'(z)} \right|_{\mat\lambda'(z) =
    \mat\lambda'_{\rm opt}(z)} = 0 \, .
\end{equation}
This Euler equation yields the optimal parameters $\mat \lambda'_{\rm
  opt}(z)$, the optimal initial-state grand potential
$\fcl\Omega_{\mat T, \mat U}[\mat \Sigma_{\mat \lambda'_{\rm opt},\mat
  U}]$, the optimal self-energy $\mat \Sigma_{\mat \lambda'_{\rm opt},
  \mat U}$, and the related one-particle Green's function
\begin{equation}
  \label{eq:gsft}
  \mat G^{\rm SFT}
  \equiv
  (\mat G_{\mat T,0}^{-1} - \mat
  \Sigma_{\mat \lambda'_{\rm opt},\mat U})^{-1} \,.
\end{equation}

The type of approximation is specified by the choice of the reference
system.  Typically, this is defined by cutting the lattice, which
underlies the original model, into disconnected clusters consisting of
a small number of sites $L_c$ each.  For Hubbard-type systems with
local interactions, this generates simple reference systems with a
small Hilbert space, which can be solved, e.g., by exact
diagonalization techniques. The resulting approximation is called the
variational-cluster approximation (VCA)
\cite{dahnken2004,potthoff2014,hofmann2013}.  A concrete example will
be given in Sec.~\ref{sec:nevca}. To enlarge the number of variational
degrees of freedom locally, a number $L_{b}$ of uncorrelated ``bath
sites'' can be coupled via a hybridization term to each of the
correlated ones.  One may formally re-derive the (nonequilibrium)
dynamical mean-field theory (DMFT)
\cite{georges1996,freericks2006,schmidt2002} by choosing a reference
system of decoupled correlated sites ($L_c=1$) hybridizing with a
continuum of bath sites ($L_b = \infty$).  The reference system is
given as a set of decoupled single-impurity Anderson models in this
case.  The VCA, the DMFT, and other approximations defined in this way
are nonperturbative by construction and can be improved
systematically by enlarging the space of variational parameters.

Self-energy functional theory in principle provides approximations
which respect the macroscopic conservation laws resulting from the
invariance of the original system under continuous U(1) and SU(2)
symmetry groups.  This is similar to the perturbatively defined
``conserving'' approaches in the sense of Baym and Kadanoff
\cite{baym1961,baym1962} which are obtained from the standard recipe
by deriving the self-energy from a truncated Luttinger-Ward
functional.  The conserving nature of the nonperturbative
approximations generated within the SFT framework is ensured by a
formal proof \cite{hofmann2013}.  This makes use of the fact that
variations of the one-particle parameters of the reference system
comprise gauge transformations of the form $\mat \lambda' \mapsto
\mat{\bar \lambda}'$ with
\begin{equation}
  \mat \varepsilon'(z) \mapsto \mat{\bar \varepsilon}'(z) = \mat
  \varepsilon'(z) - \partial_z\mat
  \chi(z)\,, 
  \qquad
  \mat T'(z) \mapsto \mat{\bar T}'(z) = e^{i\mat\chi(z)}\mat
  T'(z)e^{-i\mat\chi(z)} \, ,
  \label{eq:gaugetrans1partpar}
\end{equation}
where $\mat\varepsilon'$ denotes the (spatially) diagonal part of
$\mat\lambda'$ and $\mat T'$ its off-diagonal part, and where
$\chi_{ij,\sigma\sigma'}(z)$ is a spatially diagonal contour function.
One shows that stationarity with respect to those gauge
transformations implies that conservation laws, as expressed by
continuity equations, are proliferated from the reference system,
where they hold exactly, to the original system where they are
expressed in terms of the approximate quantities.  In fact, this is
unproblematic in practice for the case of the total particle number
and of the $z$-component of the total spin.  As discussed in Ref.\
\cite{hofmann2013}, however, respecting total-energy conservation
requires a continuum of variational degrees of freedom.  In simple
approximations, such as the VCA, one therefore has to tolerate a
violation of total-energy conservation or use this as a measure for
the quality of the approximation.


\section{Euler equation}
\label{sec:euler}

To proceed to a practical computation, one first has to note that the
trial self-energy depends on variational parameters $\mat \lambda'(z)$
which can be different on the two Keldysh branches and we therefore
distinguish $\mat\lambda'_{+}(t)$ and $\mat\lambda'_{-}(t)$ on $\mc
C_{K_+}$ and $\mc C_{K_-}$ (cf. Fig.~\ref{fig:contour}). In the end,
we are only interested in physical solutions of the Euler equation
(\ref{eq:EulerEqn}), namely solutions which satisfy
$\mat\lambda'_{+}(t)=\mat\lambda'_{-}(t)$ and hence specify a
\emph{physical} Hamiltonian. To make this more explicit, we perform a
simple rotation in parameter space and define $\mat\lambda'_{\rm
  phys}(t) = \frac{1}{2} (\mat\lambda'_{+}(t) + \mat\lambda'_{-}(t))$
and $\mat\lambda'_{\rm trans}(t) = \frac{1}{2} (\mat\lambda'_{+}(t) -
\mat\lambda'_{-}(t))$.  The physical parameter manifold is given by
$\mat\lambda'_{\rm trans}(t)=0$.  Now, it is easy to see that the
self-energy functional is trivially always stationary with respect to
physical variations when evaluated at a physical parameter set:
\begin{equation}
  \left. \frac{\delta \fcl\Omega_{\mat T,\mat U}[\mat\Sigma_{\mat \lambda', \mat U}]}{\delta \mat \lambda_{\rm phys}'(t)}  \right|_{\mat\lambda'_{\rm trans}(t) = 0}
  = 0 \: .
  \label{eq:delphys}
\end{equation}
Hence, stationarity with respect to physical variations simply places
the solution {\em on} the physical manifold.  To fix the solution {\em
  within} the physical manifold a second condition is needed, namely
stationarity with respect to the transverse variations:
\begin{equation}
  \label{eq:nontrivVar}
  \left. \frac{\delta \fcl\Omega_{\mat T,\mat U}[\mat\Sigma_{\mat \lambda', \mat U}]}{\delta \mat \lambda_{\rm trans}'(t)}  \right|_{\mat\lambda'_{\rm trans}(t) = 0}
  = 0 \: .
\end{equation}
This is in fact the central equation of the nonequilibrium SFT.
Hence, variations necessarily go off the physical manifold and thus,
opposed to the standard strategy in the equilibrium case (see e.g.\
Ref.\ \cite{balzer2010}), it is convenient to carry out the functional
derivative analytically before solving the resulting equation
numerically on the physical parameter manifold.

To compute the functional derivative in Eq.~(\ref{eq:EulerEqn}) or
(\ref{eq:nontrivVar}), we use the chain rule:
\begin{equation}
  \label{eq:EulerEqnChainRule}
  \frac{\delta \fcl\Omega_{\mat T,\mat U}[\mat\Sigma_{\mat\lambda',\mat U}]}{\delta \lambda'_{\alpha_{1}\alpha_{2}}(z)} = 
  \Tr \left( \frac{\delta \fcl\Omega_{\mat T,\mat U}[\mat\Sigma_{\mat\lambda',\mat U}]}{\delta\mat \Sigma_{\mat
        \lambda',\mat U}} \circ \frac{\delta \mat \Sigma_{\mat
        \lambda',\mat U}}{\delta \lambda'_{\alpha_{1}\alpha_{2}}(z)} \right) \,.
\end{equation}
The first factor is given by $ \beta \delta \fcl\Omega_{\mat T,\mat
  U}[\mat \Sigma_{\mat\lambda',\mat U}]/\delta \mat
\Sigma_{\mat\lambda',\mat U} = \left( \mat G_{\mat T,0}^{-1} -
  \mat\Sigma_{\mat\lambda',\mat U} \right)^{-1} - \mat
G_{\mat\lambda',\mat U}$ and can be rewritten in a more convenient
way. With the inter-cluster hopping $\mat V(z) \equiv \mat T(z) -
\mat \lambda'(z)$ we immediately have $G^{-1}_{\mat T,0}(1,2) =
G^{-1}_{\mat \lambda',0}(1,2) - \delta_{\mc C}(z_1,z_2)
V_{\alpha_1\alpha_2}(z_{2})$, see Eq.\ (\ref{eq:inverseFreeNEGF}).
Plugging this into the definition of the SFT Green's function, Eq.\
(\ref{eq:gsft}), and using Dyson's equation for the reference system,
we get
\begin{equation}
  \label{eq:CPTequation}
  \mat G^{\rm SFT} = \mat G_{\mat\lambda',\mat U} + \mat G_{\mat\lambda',\mat U} \mat V \circ \mat G^{\rm SFT} \, .
\end{equation}
This is actually the central equation of nonequilibrium
cluster-perturbation theory \cite{potthoff2011c}.  With the associated
Lippmann-Schwinger equation $\mat G^{\rm SFT} = \mat
G_{\mat\lambda',\mat U} + \mat G_{\mat\lambda',\mat U} \circ \mat
Y_{\mat\lambda',\mat T,\mat U} \circ \mat G_{\mat\lambda',\mat U}$,
where $\mat Y_{\mat\lambda',\mat T,\mat U}(z_{1},z_{2}) = \mat
V(z_{1}) \delta_{\mc C}(z_{1},z_{2}) + \mat V(z_{1}) \mat G^{\rm
  SFT}(z_{1},z_{2}) \mat V(z_{2})$ is the corresponding $T$-matrix, we
eventually obtain
\begin{equation}
  \label{eq:first}
  \frac{\delta \fcl \Omega_{\mat T,\mat U}[\mat\Sigma_{\mat \lambda',\mat U}]}{\delta\mat \Sigma_{\mat \lambda',\mat U}} = \frac{1}{\beta} \mat G_{\mat\lambda',\mat U} \circ \mat
  Y_{\mat\lambda',\mat T,\mat U} \circ \mat G_{\mat\lambda',\mat U} \:
\end{equation}
for the first factor in Eq.\ (\ref{eq:EulerEqnChainRule}).  To
evaluate the second factor, we write $\mat\Sigma_{\mat\lambda',\mat U}
= \mat G_{\mat\lambda',0}^{-1} - \mat G_{\mat\lambda',\mat U}^{-1}$.
Apart from the $\mat \lambda'$-dependence of the inverse free Green's
function, see Eq.~(\ref{eq:inverseFreeNEGF}), we have to consider the
derivative of the inverse of the full Green's function: From the
identity $\delta(\mat G_{\mat\lambda',\mat U}\circ\mat
G_{\mat\lambda',\mat U}^{-1})/\delta\mat\lambda' = 0$ we deduce
$\delta \mat G_{\mat\lambda',\mat U}^{-1}/\delta\mat\lambda' = - \mat
G_{\mat\lambda',\mat U}^{-1}\circ\delta \mat G_{\mat\lambda',\mat
  U}/\delta\mat\lambda'\circ\mat G_{\mat\lambda',\mat U}^{-1}$.
Finally, making use of the properties of the time ordering operator
and the exponential function, a straightforward calculation yields:
\begin{equation}
  \label{eq:dGdlambda}
  \frac{\delta G_{\mat\lambda',\mat U}(3,4)}{\delta
    \lambda'_{\alpha_1\alpha_2}(z_1)} = G_{\mat\lambda',\mat U}(3,4)
  \left. G_{\mat\lambda',\mat U}(2,1^+) \right|_{z_2=z_1}  -
  \left. G^{(2)}_{\mat\lambda',\mat U}(3,2,1^+,4) \right|_{z_2=z_1} \,,
\end{equation}
where $\mat G_{\mat\lambda',\mat U}^{(2)}$ is the two-particle Green's
function of the reference system.  Collecting the results we find:
\begin{equation}
  \label{eq:K0}
  -\beta \frac{\delta \fcl \Omega_{\mat T,\mat U}[\mat\Sigma_{\mat
      \lambda',\mat U}]}{\delta \lambda'_{\alpha_1\alpha_2}(z_1)} =
  \dint d3d4 \, Y_{\mat \lambda',\mat T,\mat
    U}(4,3) \left. L_{\mat\lambda',\mat U} (3,2,1^+,4)
  \right|_{z_2=z_1} =: K^{(0)}[\mat\lambda']_{\alpha_2\alpha_1}(z_1) \,,
\end{equation}
where $L_{\mat\lambda',\mat U}(1,2,3,4) = G_{\mat\lambda',\mat
  U}(2,4)G_{\mat\lambda',\mat U}(1,3) - G_{\mat\lambda',\mat
  U}(1,4)G_{\mat\lambda',\mat U}(2,3) + G_{\mat\lambda',\mat
  U}^{(2)}(1,2,3,4)$ is the two-particle (four-point) vertex function
with external legs.  Therewith, we have the Euler equation of the
nonequilibrium SFT:
\begin{equation}
  \label{eq:sfteuler}
  \left.\mat K^{(0)}[\mat\lambda'](t)\right|_{\mat\lambda'=\mat\lambda'_{\rm opt}} = 0\,,
\end{equation}
which will be the starting point for the numerical determination of
the optimal parameters. Note that those parameters depend on the real
time variable $t$ only rather than on the contour variable $z$,
because after the transverse variations have been carried out
analytically we can restrict the search for optimal parameters to the
physical manifold, see Eq.\ (\ref{eq:nontrivVar}).


\section{Numerical implementation}
\label{sec:numer-impl}

\begin{figure}[t]
  \includegraphics[width=.48\textwidth]{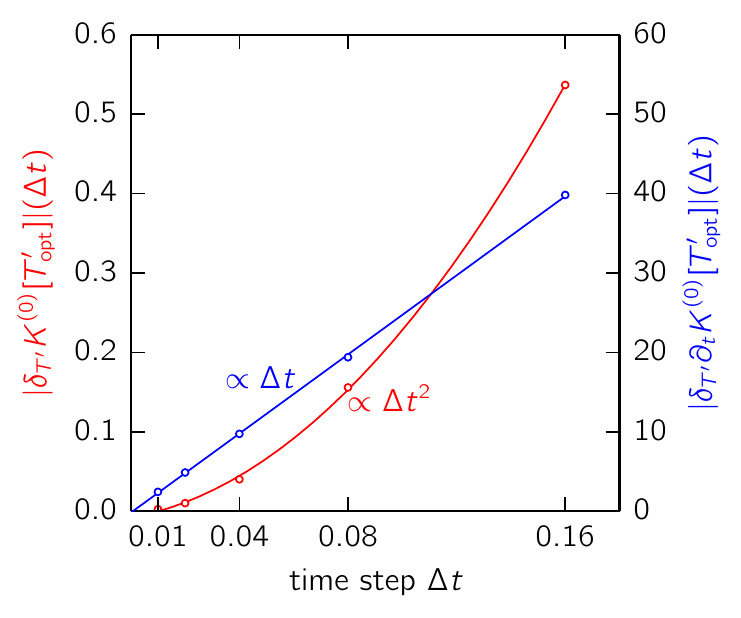}\hspace{.04\textwidth}%
  \includegraphics[width=.48\textwidth]{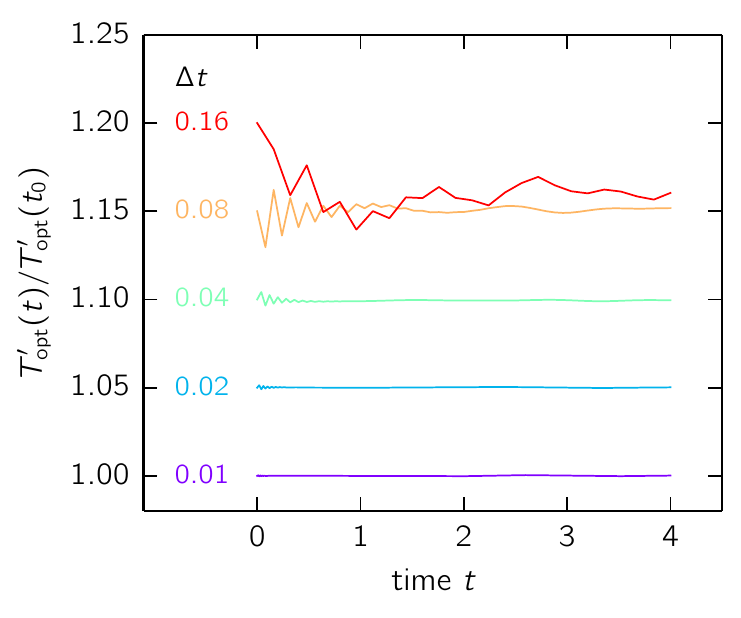}
  \caption{ Test of the numerical implementation for the
    time-propapagation of a system in equilibrium.  The system is a
    one-dimensional chain of $100$ sites at $\beta = 6$ and $U=4$. As
    a reference system a two-site cluster is used. \emph{Left}:
    $\Delta t$ dependence of the Jacobian of $\mat
    K^{(0)}[\mat\lambda'{]}(t)$ (red, left ordinate axis) and of
    $\partial_t\mat K^{(0)}[\mat\lambda'{]}(t)$ (blue, right ordinate
    axis). Note, that the ordinate axes' scales differ by two orders
    of magnitude.  Results are independent of the time
    $t$. \emph{Right}: Time dependence of the optimal parameters. VCA
    results obtained as the roots of $\partial_t\mat
    K^{(0)}[\mat\lambda'{]}(t) = 0$ [Eq.~(\ref{eq:dK0dt})] for
    different $\Delta t$.  The ordinate axis refers to the result for
    $\Delta t = 0.01$; results for larger $\Delta t$ are constantly
    shifted by multiples of $0.05$.}
  \label{fig:K_dtscaling}
\end{figure}

With the \emph{time-dependent} Euler equation~(\ref{eq:sfteuler}) we
are facing a profound root-finding problem of formally infinite
dimensions.  However, the SFT variational principle is inherently
causal which allows us to determine the optimal parameters at a given
time on a discrete time grid, without affecting the results at
previous grid times, and then proceed to the next time.  This
causality is most easily seen when discussing Eq.~(\ref{eq:K0}): The
integrals over $z_3$ and $z_4$ extend over the entire contour $\mc C$
but can be confined to times which are (physically) earlier than
$t_{\rm max}=t_1$ (cf.\ Fig.~\ref{fig:contour}).  If one or both times
are located beyond $t_1$, the later time can be shifted from the upper
to the lower branch (or vice versa) without altering the contour
ordering, and the respective contributions to the integral cancel.
Thus, at time $t_{1}$, all quantities in the Euler
equation~(\ref{eq:sfteuler}) and therewith the parameter $\mat
\lambda'_{\rm opt}(t_{1})$ itself depend on parameters at earlier
times only.

Unfortunately, a straightforward numerical solution of
Eq.~(\ref{eq:sfteuler}) turns out to be impossible.  The optimal
parameters, determined by standard root-finding techniques, quickly
accumulate a large error after a few time steps only such that a
reasonable solution cannot be found in this way.  Generally, for a
functional depending parametrically on time, which is evaluated on a
finite time grid and the parameters of which are varied only at the
very last instant of time, we expect the explicit dependence of the
variation of the functional on the time step $\Delta t$ to scale as
$(\Delta t)^n$ with $n\geq 1$, since for an infinitesimally fine grid
variations at a single instant of time would reduce to variations on a
null set, and thus the variation of the functional must vanish.

In the present case, if one aims to solve Eq.~\eqref{eq:sfteuler} at a
given time $t$, keeping the solution $\mat\lambda'_{\rm opt}$ fixed at
earlier times, the variation $\delta_{\mat\lambda'(t)}\mat
K^{(0)}[\mat\lambda'_{\rm opt}](t)$ of the parameters at the last
time-step will scale as $(\Delta t)^2$, if the whole implementation is
based on an equidistant time-grid $\Delta t$.  This is verified
numerically in Fig.~\ref{fig:K_dtscaling} (left) for a special case
(described in the figure caption). This scaling is independent of the
quadrature rules used in solving the time integrals (if the accuracy
of those scales as $(\Delta t)^2$ or better).  Hence the observed
$(\Delta t)^2$ scaling is in fact due to the Jacobian
$\delta_{\mat\lambda'(t)}\mat K^{(0)}[\mat\lambda'_{\rm opt}](t)$
itself and turned out to be detrimental for the stability of the
algorithm for practically reasonable choices of $\Delta t$.

Fortunately, this problem can be overcome by requiring the \emph{time
  derivative} of $\mat K^{(0)}[\mat\lambda'](t)$ [see
Eq.~(\ref{eq:K0})] to vanish for all times $t>t_0$ and by fixing the
initial conditions at $t_0$ by Eq.~(\ref{eq:sfteuler}).  Hence,
instead of solving the Euler equation (\ref{eq:sfteuler}) for all
times, we rather consider
\begin{subequations}
  \label{eq:DiffEulerEqn}
  \begin{align}
    \left.\mat
      K^{(0)}[\mat\lambda'](t)\right|_{\mat\lambda'=\mat\lambda'_{\rm
        opt}} &= 0 \,, \qquad \text{for } t=
    t_0\,,  \label{eq:DiffEulerEqn_eq} \\
    \left.\partial_t \mat
      K^{(0)}[\mat\lambda'](t)\right|_{\mat\lambda'=\mat\lambda'_{\rm
        opt}} &= 0 \,, \qquad \text{for } t>
    t_0\,. \label{eq:DiffEulerEqn_ne}
  \end{align}
\end{subequations}
As can be seen in Fig.~\ref{fig:K_dtscaling} (left), the dependence on
$\Delta t$ is only linear in this case.  This greatly improves the
accuracy of the parameter optimization and allows us to trace the
optimal variational parameters as a function of time.  As a simple
numerical check, one may consider the {\em equilibrium} problem.  The
expected time {\em in}dependence of the optimal variational parameters
is indeed found for sufficiently small $\Delta t$, see
Fig.~\ref{fig:K_dtscaling} (right).

According to Eq.~(\ref{eq:K0}), the time derivative $\partial_t \mat
K^{(0)}[\mat\lambda'](t)$ can be obtained from the corresponding
equations of motion for the four-point vertex function $\mat L_{\mat
  \lambda',\mat U}$.  There are two qualitatively different
dependencies on $t$: The first one results from the boundary terms due
to the time-ordering operator comprised by all Green's functions.
Contributions from this one cancel out when taking the time
derivative.  The second one is the explicit dependence of the
annihilator and the creator on the external time $t_1$.  This is
governed by the Heisenberg equation of motion.  Commuting the
operators with the one-particle part of the Hamiltonian simply results
in matrix products with $\mat \lambda'$ while commuting with the
interacting part gives rise to higher-order products of annihilators
and creators which we denote by $\psi$ or $\psi^{\dagger}$,
respectively: $\com[\ac{}(1),\op H'_1(1)] \equiv \psi(1)$ and
$\com[\op H'_1(1),\cc{}(1)] \equiv \psi^\dagger(1)$.  Thus, the time
derivative of $\mat K^{(0)}[\mat\lambda'](t)$ acquires the form:
\begin{equation}
  \partial_t \mat K^{(0)}[\mat\lambda'](t) = \com[\mat K^{(0)}[\mat\lambda'](t),\mat \lambda'(t)] + \mat
  K^{(1)}[\mat\lambda'](t) \,, \label{eq:dK0dt}
\end{equation}
where we have used Eq.~(\ref{eq:K0}) and where
\begin{equation}
  \label{eq:K1}
  K^{(1)}[\mat\lambda']_{\alpha_2\alpha_1}(z_1) = 
  \dint d3d4\left. Y_{\mat \lambda',\mat T,\mat U}(4,3)
    M_{\mat\lambda',\mat U} (3,2,1^+,4) \right|_{z_2=z_1} \,,
\end{equation}
with
\begin{equation}
  M_{\mat\lambda',\mat U} (3,2,1^+,4) = L_{\mat\lambda',\mat U} (3,2,1^+_{\psi},4) - L_{\mat\lambda',\mat
    U} (3,2_\psi,1^+,4) \,.
\end{equation}
The subscript $\psi$ indicates that $\ac{}$ and $\cc{}$ are replaced
by $\psi$ and $\psi^\dagger$ in the respective correlation function.
For example, $i G_{\mat \lambda',\mat U}(1_\psi,2) = \ev< \ord_{\mc C}
\psi(1)\cc{}(2) >_{\mat \lambda', \mat U}$.

\begin{figure}[t]
  \centering
  \includegraphics[width=.8\textwidth]{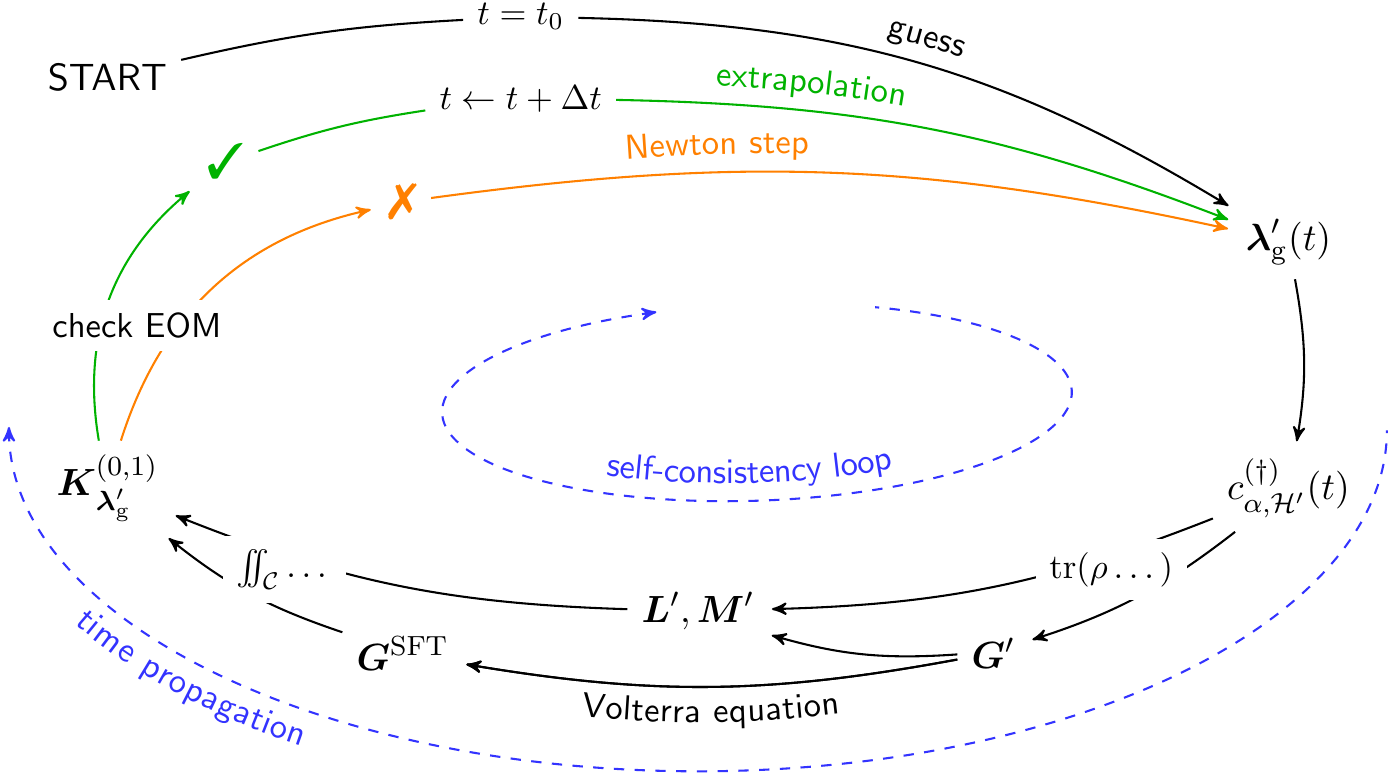}
  \caption{Sketch of the propagation algorithm for the optimal
    parameters. See text for discussion.}
  \label{fig:propscheme}
\end{figure}

The algorithm for the numerical implementation of the VCA in the
general nonequilibrium case [Eq.~(\ref{eq:DiffEulerEqn})] is sketched
in Fig.~\ref{fig:propscheme}. The following steps are performed at
each time step:
\begin{itemize}
\item[(i)] At a given time $t$ and for certain guessed parameters
  $\mat\lambda'_{\rm g}(t)$, we propagate $\cc{\alpha}(t)$ and
  $\ac{\alpha}(t)$ for all relevant orbitals $\alpha$ by a time step
  $\Delta t$ and store their respective representations in the
  occupation-number basis.  For small cluster sizes, this is
  straightforward.  Therewith, arbitrary correlation functions can be
  calculated for arbitrary times on the contour up to the time $t$.
  While the single particle Green's function $\mat G'$ can be updated
  from time step to time step and kept in the storage, the higher
  correlation functions $\mat L'$ and $\mat M'$ have to be
  recalculated on-the-fly for \emph{any} time step and after any
  update of $\mat\lambda'_{\rm g}(t)$, because of the external time
  $t$.  Symmetries can be exploited to reduce the actual number of
  elements that have to be calculated.
\item[(ii)] The SFT Green's function $\mat G^{\rm SFT}$ is obtained
  from the CPT equation (\ref{eq:CPTequation}), which can be cast in
  the form of a Volterra equation of second kind: $(\mat 1 - \mat
  G_{\mat\lambda',\mat U} \mat V )\circ\mat G^{\rm SFT} = \mat
  G_{\mat\lambda',\mat U}$.  The latter is solved up to time $t$ by
  standard techniques
  (cf. Refs.~\cite{eckstein2009,eckstein2010b,brunner1986}).  Here, we
  exploit the translational symmetries of $\mat V$ and of $\mat G^{\rm
    SFT}$ by Fourier transformation with respect to the super-lattice.
\item[(iii)] The correlation functions $\mat L'$, $\mat M'$ and $\mat
  G^{\rm SFT}$, are then used to calculate both $\mat
  K^{(0)}[\mat\lambda'](t)$ and $\mat K^{(1)}[\mat\lambda'](t)$ via
  Eqs.~(\ref{eq:K0}) and (\ref{eq:K1}).  For the initial time $t_0$,
  only those integrations contribute where both times are on the
  Matsubara branch.  For any later time, the mixed and Keldysh
  integrations have to be carried out, too.  Results for the integrals
  from earlier time steps cannot be recycled, due to the external time
  $t$ appearing in the correlation functions $\mat L'$ and $\mat M'$.
  For both, the integrations involved in the Euler
  equation~(\ref{eq:DiffEulerEqn}) as well as in the Volterra
  equation, we use high-order integration schemes, like the
  Newton-Cotes rules or the Gregory rules
  \cite{brunner1986,press2007}, to allow for large $\Delta \tau$ and
  large $\Delta t$ steps \cite{CFET}.
\item[(iv)] Generally, the initial guess $\mat \lambda'_{\rm g}(t)$
  will not be a root of Eqs.~(\ref{eq:DiffEulerEqn}), and must
  therefore be updated.  The standard way of doing this, is to apply
  Newton's method.  This, however, requires the knowledge of the
  (inverse of the) functional's Jacobian
  $\delta_{\mat\lambda'(t_0)}\mat K^{(0)}[\mat\lambda'](t_0)$ or
  $\delta_{\mat\lambda'(t)}\partial_t\mat K^{(0)}[\mat\lambda'](t)$
  respectively.  Both, the implementation of an analytical expression
  for the Jacobian or a direct numerical evaluation via finite
  difference methods are rather costly options and thus not feasible.
  We thus employ Broyden's method \cite{kelley1987,broyden1965}, which
  provides increasingly improved updates for the (inverse of the)
  Jacobian during the course of the Newton iteration, starting from an
  initial guess for both, the root as well as the Jacobian itself.
  However, at the initial time we evaluate the Jacobian numerically by
  finite differences at some guessed parameter.  This improves the
  success and the speed of the method significantly.  At later times,
  we extrapolate both the optimal parameter as well as the inverse
  Jacobian of $\partial_t \mat K^{(0)}[\mat\lambda'](t)$ to again
  start the Broyden iteration from an accurate initial guess - with
  this, we reliably find the roots of the respective functional after
  only one to three iterations per time step.
\end{itemize}

Finally, we would like to add an important remark regarding the
solution of Eq.\ (\ref{eq:DiffEulerEqn_ne}). As Eq.\ (\ref{eq:dK0dt})
contains $\mat\lambda'(t)$ explicitly, a first naive approach would be
to guess an ``optimal'' value $\mat\lambda'_{\rm g}(t)$, calculate
$\mat K^{(0)}[\mat\lambda'_{\rm g}](t)$ and $\mat
K^{(1)}[\mat\lambda'_{\rm g}](t)$, plug these into
Eq.~(\ref{eq:dK0dt}), solve the resulting equation $ 0 = \com[\mat
K^{(0)}[\mat\lambda'_{\rm g}](t),\mat \lambda'(t)] + \mat
K^{(1)}[\mat\lambda'_{\rm g}](t)$ for a new $\mat\lambda'(t)$, update
both functionals and iterate this procedure until convergence.
However, there are two complications.  First, close to the optimal
point, $\mat K^{(0)}[\mat\lambda'](t)$, and hence also $\mat
K^{(1)}[\mat\lambda'](t)$ must vanish and therefore solving
Eq.~(\ref{eq:dK0dt}) for $\mat\lambda'(t)$ will get more and more
unstable.  Second, for a guessed parameter it is unlikely that the
corresponding functionals will be compatible with
Eq.~(\ref{eq:dK0dt}), i.e., that a solution is existing at all.  For
this to be the case, e.g., $\mat K^{(1)}[\mat\lambda'](t)$ has to be
trace-less, since it otherwise could not equal a commutator of two
other matrices. Indeed, in our numerics this requirement is not met in
general. More generally, Eqs.~(\ref{eq:dK0dt}) and
(\ref{eq:DiffEulerEqn}) can be understood as a special case of
Sylvester's equation, namely $\mat A \mat X - \mat X \mat B = \mat C$,
which has a unique solution for any matrix $\mat C$, if and only if
the matrices $\mat A$ and $\mat B$ have distinct spectra (see
Refs.~\cite{sylvester1884,rosenblum1956,bhatia1997}). In our case,
this is clearly not the case, which is why we cannot expect a unique
solution (or any solution at all), for some $\mat
K^{(1)}[\mat\lambda'](t)$ as obtained for an arbitrarily guessed
optimal parameter. A way out would be to not solve
Eq.~(\ref{eq:DiffEulerEqn}) directly, but minimize the norm of the
left-hand side, which then, in any case, again suffers from the first
complication.


\section{Application of the nonequilibrium variational-cluster
  approach}
\label{sec:nevca}

In the following, two concrete examples for the application of the
nonequilibrium variational-cluster approach are presented to discuss
some of its characteristics in detail.  We consider the
one-dimensional Hubbard model at half-filling and at different
interaction strengths and study quenches (or fast ramps) of the
hopping parameters.  Both examples have been in the focus of recent
experiments where the redistribution of antiferromagnetic correlations
between different bonds and for different ramp times \cite{greif2015}
as well as the topological properties of Bloch bands in optical
lattices \cite{atala2013} for the uncorrelated variant of the model
\cite{su1979,rice1982} were studied.

\begin{figure}[b]
  \subfloat[dimerized lattice $\to$ homogeneous lattice]
  {\label{fig:VCAramp_inifin}\includegraphics[width=.45\textwidth]{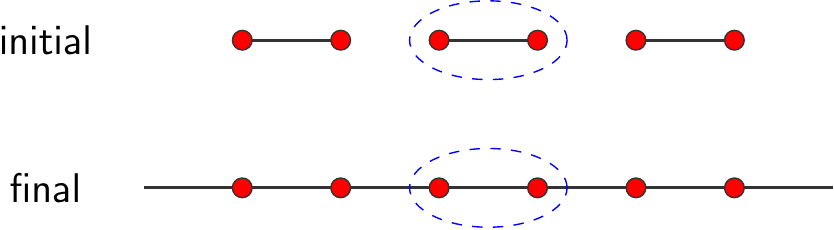}}\hspace{.1\textwidth}%
  \subfloat[dimerization change]
  {\label{fig:VCAflip_inifin}\includegraphics[width=.45\textwidth]{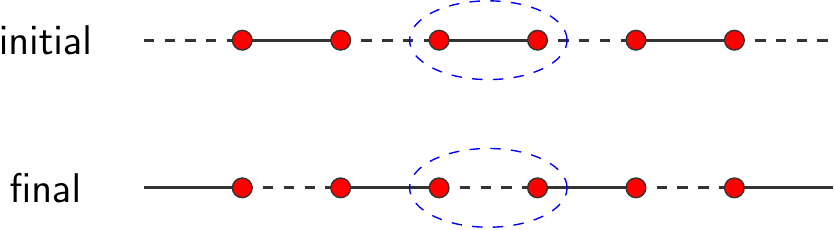}}
  \caption{ Illustration of initial and final states for both (a) the
    quench from a dimerized configuration to a homogeneous lattice as
    well as (b) change of the dimerization.  Black solid lines
    indicate a nearest-neighbor intra- or inter-cluster hopping,
    $T_{\rm intra}=1$ or $T_{\rm inter}=1$.  Black dashed lines:
    $T_{\rm inter}=0.2$. Correlated sites are represented by red
    filled dots.  The reference system used for the VCA calculations
    is indicated by a representative two-site reference cluster
    highlighted as the blue dashed ellipse.}
  \label{fig:VCAinifin}
\end{figure}

Figure~\ref{fig:VCAinifin} provides an illustration of the initial and
of the final states as well as of the reference systems: In both
cases, the system is initially in the thermal state of a dimerized
Hubbard model specified by some inverse temperature $\beta$.  In case
\subref{fig:VCAramp_inifin}, this state is generated by an initial
Hamiltonian which consists of decoupled two-site clusters.  Hence, the
initial state is a simple valence-bond state with nearest-neighbor
correlations and reduced translational symmetries.  The intra-cluster
hopping $T_{\rm intra} = 1$ fixes energy and time units.  In case
\subref{fig:VCAflip_inifin}, the clusters with $T_{\rm intra} = 1$ are
weakly coupled by an inter-cluster hopping $T_{\rm inter} =0.2$.

In case \subref{fig:VCAramp_inifin}, the final-state dynamics after
the quench is governed by the full Hubbard Hamiltonian where the
initially disconnected clusters are linked by a final inter-cluster
hopping $T_{\rm inter} = 1$.  This is a highly nontrivial example
where the system should build up longer-ranged nonlocal correlations
and entanglement in the course of time.  In case
\subref{fig:VCAflip_inifin} the final-state Hamiltonian is basically
the same as the Hamiltonian specifying the initial state but with the
important difference that the nearest-neighbor hopping $T_{\rm
  inter}=1$ now connects clusters that are shifted by one lattice
constant.  This example is also highly nontrivial as it corresponds
to a sudden switch between Hamiltonians describing states with
well-developed but incompatible valence bonds where the entanglement
and the spin correlations must reorganize between two different local
situations.

The VCA, as a cluster mean-field approximation, can only partly cover
the expected final-state dynamics.  Clearly, the quality of the
approximation decisively depends on the size of the cluster used in
the reference system.  Note that in both cases one in principle needs
clusters of infinite size to recover the exact solution.  Here,
however, our main intention is to discuss some numerical issues and to
demonstrate that the VCA can be implemented successfully and yields
reasonable results (which can in principle be improved systematically
by going to larger cluster sizes). To this end, we have choosen a
simple reference system, namely a system of disconnected clusters
consisting of two sites each, indicated by the blue dashed lines in
Fig.~\ref{fig:VCAinifin}.  No additional bath degrees of freedom have
been added.  Thus, the only possible variational parameters are the
intra-cluster hopping $T'$ and the on-site energies $\varepsilon'_i$
in the reference system. The latter are fixed by the symmetries of the
model at half-filling and, therefore, only the intra-cluster hopping
$T'$ is left for optimization.

We have tested the computer code in several trivial limits.
Furthermore, for the equilibrium case our data for different $U$ and
$\beta$ and for a \emph{homogeneous} lattice are fully consistent with
those obtained previously for $\beta\ra\infty$ in Ref.\
\cite{balzer2008} using a completely different algorithm.

Calculations for the systems sketched in Fig.~\ref{fig:VCAinifin} have
been performed at inverse temperature $\beta = 10$ which, on the level
of the approximation employed, is already representative of the
zero-temperature limit as has been checked by varying $\beta$.  Using
fifth-order integration schemes on the Matsubara branch, converged
results are obtained with $\Delta \tau = 0.1$.  On the Keldysh branch
we are limited to the trapezoidal rule only since the implementation
of higher-order schemes for the evaluation of the time-propagation
operator gives rise to numerical instabilities (see
Sec.~\ref{sec:numer-impl} and comment~\cite{CFET}). Converged results
are obtained for a time step of $\Delta t = 0.02$, and a maximum
propagation time of $t_{\max} = 10$ is easily achieved with a desktop
computer.  To get converged results with respect to the spatial
extension of the one-dimensional lattice, it is sufficient to consider
systems with $L=100$ sites (using periodic boundary conditions).
Rather than a sudden parameter quench we assume a finite (but short)
ramp within a time interval $\Delta t_{\rm ramp} = 0.5$, using a ramp
profile $r(t) = (1-\cos(\pi t/\Delta t_{\rm ramp}))/2$ with continuous
first-order derivatives at the joints at $t=0$ and $t=\Delta t_{\rm
  ramp}$.  This has been recognized to stabilize the algorithm
significantly.  It is also numerically advantageous to hold the
original system in its initial equilibrium state for a few time steps,
before starting with the ramp at $t=t_0=0$ to build up the desired
integration order.

\begin{figure}[t]
  \subfloat[dimerized lattice $\to$ homogeneous
  lattice]{\label{fig:VCAramp}\includegraphics[width=.48\textwidth]{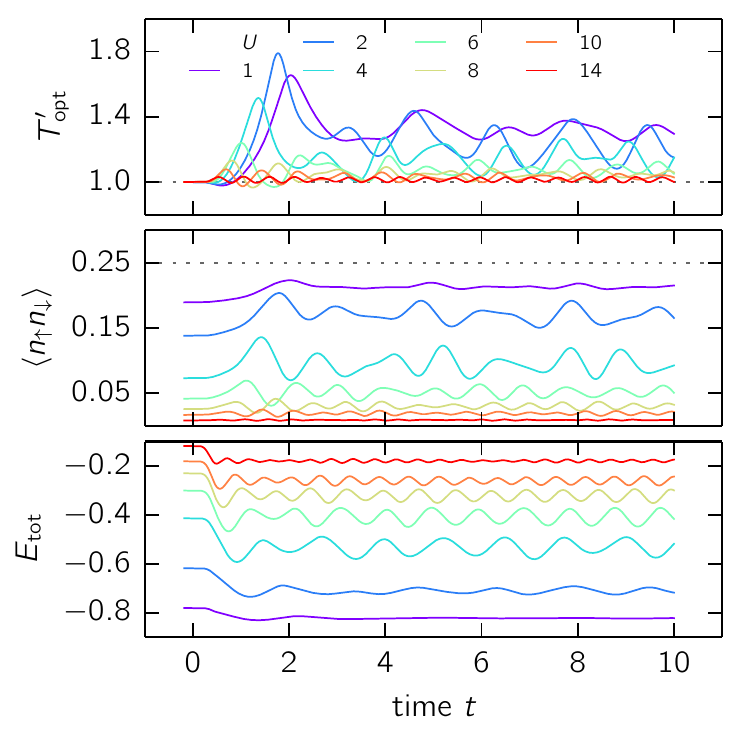}}\hspace{.04\textwidth}%
  \subfloat[dimerization
  change]{\label{fig:VCAflip}\includegraphics[width=.48\textwidth]{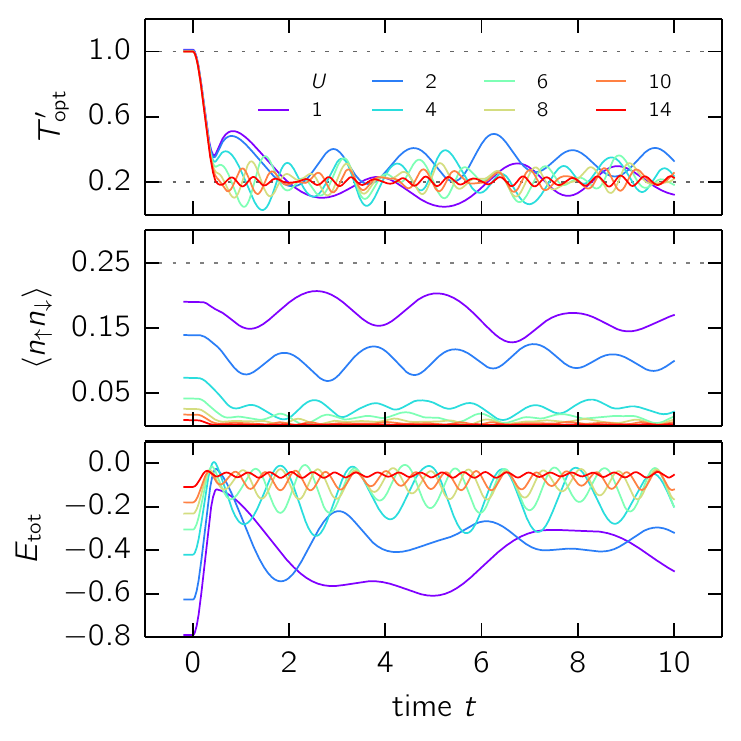}}
  \caption{ Time dependencies of the optimal hopping $T'_{\rm opt}$ of
    the two-site reference cluster, the double occupancy and the total
    energy $E_{\rm tot}(t) = \langle H(t) \rangle$ for different $U$ and for the two different ramps of the
    hopping parameters of the original system (see
    Fig.~\ref{fig:VCAinifin}, a and b).}
  \label{fig:VCA}
\end{figure}

Results are shown in Fig.~\ref{fig:VCA}.  We first discuss case
\subref{fig:VCAramp_inifin}.  For $t=0$, the optimal value of the
variational parameter is found as $T_{\rm opt}'=T_{\rm intra}=1$
(dotted line).  In fact, this had to be expected since, for the given
problem, the self-energy of the reference system equals the full
self-energy [cf.\ Eq.~(\ref{eq:gsft})] and thus the VCA is exact in
the initial state.  This represents another nontrivial check of our
algorithm.

For $t>0$ we find that $T_{\rm opt}'$ becomes time-dependent, i.e.,
the reference system adjusts itself to the parameter ramp in a
time-dependent way to optimally describe the dynamics of the original
system.  In the limit of infinitely large clusters where VCA formally
becomes exact, one would expect $T_{\rm opt}'(t)$ to become constant
after a certain relaxation time.  As seen in the figure, this
relaxation of the optimal variational parameter, and also of the
double occupancy, takes place on a short time scale given by one or
two inverse hoppings.  With a reference cluster of two sites only,
however, some finite-size effects must be tolerated.  These show up
indeed as oscillations of $T_{\rm opt}'(t)$ around an average value
after the relaxation process.  For weak $U$, where the physics is
governed by the hopping part of the Hamiltonian, this average is by
about 30\% higher than the initial value $T_{\rm intra}=1$.  With
increasing $U$, the final average value of $T_{\rm opt}'(t)$ is seen
to decrease and approaches unity for $U\to \infty$.  This is plausible
since for weak $U$ the fermion system is more itinerant and thus an
increased hopping parameter in the reference system is necessary to
(at least partially) compensate for the missing inter-cluster hopping
in the reference system while for strong $U$ this is less important as
the physics is more local.

Similar arguments can be used to explain the time dependence of the
double occupancy.  For $t=0$, there is a strong $U$ dependence of
$\langle n_{\uparrow} n_{\downarrow} \rangle$, which is exactly
reproduced by the VCA.  For $t>0$, the double occupancy quickly
relaxes to a higher value (apart from significant finite-size
oscillations) as the system becomes more itinerant due to the
additional physical inter-cluster hopping.  The effect is strongest
for weak $U$.

For case \subref{fig:VCAflip_inifin}, finite-size effects are
generally somewhat stronger, see Fig.~\ref{fig:VCA} (right).  Still
the main trends are clear and plausible: As the inter-cluster hopping
is weak, the initial values of $T_{\rm opt}'$ and of $\langle
n_{\uparrow} n_{\downarrow} \rangle$ are close to those of case
\subref{fig:VCAramp_inifin}. For $t>0$, the relaxation process takes
place on essentially the same time scale of one or two inverse
hoppings.  However, one now expects that the optimal intra-cluster
hopping adjusts to a value close to $T_{\rm inter}=0.2$, i.e., close
to the physical hopping parameter (see final state in
Fig.~\ref{fig:VCAinifin}\subref{fig:VCAflip_inifin}).  This is in fact
seen (dotted line in Fig.~\ref{fig:VCA}, right).  The decrease of the
final average value of the double occupancy with increasing $U$ is
understood in the same way as in case \subref{fig:VCAramp_inifin}.

Note that the geometrical structure of the reference system is the
same for the initial and the final state, see
Fig.~\ref{fig:VCAinifin}. Given this, we expect that the description
of the initial state is better than that of the final state in the
case of the dimerization change.  We have also performed calculations
using a reference system that is shifted by one lattice constant for
both, the initial and the final state.  In this case, one expects that
the VCA description of the final-state dynamics is more accurate than
for the initial state.  The calculations (not shown) yield unphysical
results in this case with a diverging optimal intra-cluster hopping
after one to two inverse hoppings, depending somewhat on $U$.  This
demonstrates the crucial importance of an accurate description of the
initial state for the subsequent real-time dynamics.  A sudden switch
of the geometrical structure of the reference system at $t=0$, which
follows the dimerization change of the original system, would be a
legitimate choice and would result in a superior approximation.  This,
however, requires a substantially higher numerical effort as more than
a single two-site cluster must be considered as a building block in
the calculation which is beyond the scope of the present work.

In both cases, \subref{fig:VCAramp_inifin} and
\subref{fig:VCAflip_inifin}, there is a fairly good conservation of
the total energy right after the ramp (recall that $\Delta t_{\rm
  ramp} = 0.5$) with some remaining finite-size oscillations but no
long-time drift.  Note that, {\em a priori}, this could not have been
expected as a matter of course, since strict energy conservation
within SFT can only be ensured by variations \emph{nonlocal} in time,
which would correspond to the optimization of infinitely many
bath-degrees of freedom (see Ref.~\cite{hofmann2013} for a detailed
discussion).

In case \subref{fig:VCAramp_inifin} and for all interaction strengths,
we find that the total energy decreases after the ramp (see
Fig.~\ref{fig:VCAramp}).  This implies that the increase of the total
energy due to the heating of the system during the ramp is
overcompensated by the energy decrease that is induced by the coupling
of the different isolated clusters via $\op H_{T_{\rm inter},0}$ and
the corresponding lowering of the kinetic energy.  In case
\subref{fig:VCAflip_inifin} the total energy increases for all $U$
after the ramp (see Fig.~\ref{fig:VCAflip}).  Since the initial and
the final Hamiltonians are identical apart from a translation by one
lattice constant, they have the same ground-state energies.  The
observed increase of the total energy must therefore be exclusively
due to the heating of the system during the ramp.


\section{Summary}
\label{sec:summary-outlook}

The self-energy functional theory (SFT) provides a unifying framework
for different types of cluster and impurity approximations and has
recently been extended to nonequilibrium cases \cite{hofmann2013}.
In the present work, we have discussed its numerical implementation in
detail. An important observation is that a straightforward solution of
the central SFT Euler equation is impossible in practice as it suffers
from numerical instabilities which could traced back to an inherent
quadratic dependence of the Jacobian on the time step $\Delta t$.
Fortunately, the Euler equation can equivalently be replaced by its
time derivative (and the appropriate initial condition).  The
corresponding Jacobian shows a numerically much more favorable
linear-in-$\Delta t$ scaling.  By using high-order integration schemes
and Broyden's method we have put forward and have implemented a stable
propagation algorithm for the optimal parameters of the reference
system.

As a concrete example, the variational cluster approach (VCA) has been
applied to study the dynamics initiated by two different ramps of the
hopping parameters of an initially dimerized one-dimensional Hubbard
model.  The time-evolution of the optimal parameters, the double
occupancy and the total energy has been studied at different
interaction strengths.  These calculations have been performed with a 
two-site reference cluster and demonstrate that plausible and
consistent results can be obtained in fact.
Note that, as compared to plain CPT, the strengths of the VCA are expected to show up when considering situations involving (dynamical) phase transitions or in the possibility to construct approximations involving bath degrees of freedom. 
This, however, is beyond the scope of the present paper. 

Despite the simple two-site reference system used here, the resulting real-time dynamics is in fact completely different from a mere superposition of oscillations with frequencies that are characteristic for the finite reference system.  
Namely, the variational embedding of the cluster rather allows to describe the
relaxation of the system to a new stationary final state.  Depending
on the system, on the type of process studied and on the model
parameters, however, the final state does show some unphysical
oscillations which are caused by the small size of the reference
system and which must be tolerated at the given approximation level.
The SFT framework allows to systematically improve the approximation
by consideration of larger clusters, clusters with more variational
parameters or by attaching uncorrelated bath sites.  We expect that
the remaining finite-size oscillations can be systematically
controlled in this way.

In practice, calculations for larger clusters are computationally expensive. 
Since the exact intra-cluster 4-point correlation function $L(3,2,1^{+},4)$ is required, the practically accessible cluster size is {\em eventually} limited by the exponential growth of the cluster Hilbert-space dimension.
However, for small clusters the computational bottleneck of the algorithm actually consists in the necessity to repeatedly calculate the correlation functions $\mat K^{(0)}[\mat\lambda'](t)$ (and $\mat K^{(1)}[\mat\lambda'](t)$) at each instant of time and for each Newton step via a double contour integration and a double sum over the intra-cluster orbitals (see Eqs.~(\ref{eq:K0}) and (\ref{eq:K1})).
With the present implementation, clusters with 6 or more sites are clearly beyond our capabilities.

Approximations generated within the SFT framework do respect
macroscopic conservation laws in general \cite{hofmann2013}.
Regarding total-energy conservation, however, this would require the
optimization of parameters which are nonlocal in time, i.e., the
allocation of infinitely many bath sites. Since this has not been
considered here, we also have to tolerate superimposed finite-size
oscillations on the total energy.  Apart from those, however,
conservation of the total energy is described fairly well, in
particular, there is no long-time drift.  However, to answer the
question of whether or not the stationary final state is a thermal
state that can be described by the temperature corresponding to the
total energy after the parameter quench or ramp, in fact requires
improved approximations and larger reference systems.


\ack We would like to thank K.\ Balzer for providing computer code for
setting up the many-body basis and for CPT reference calculations as
well as for helpful discussions. We would like to thank D.\ Gole\v{z},
C.\ Gramsch, H.\ Strand for stimulating discussions. Support of this
work by the Deutsche Forschungsgemeinschaft within the
Sonderforschungsbereich 925 (projects B5 and B4) is gratefully
acknowledged.


\section*{References}

\providecommand{\newblock}{}

\end{document}